\documentclass[conference]{IEEEtran}
\IEEEoverridecommandlockouts
\usepackage{cite}
\usepackage{amsmath,amssymb,amsfonts}
\usepackage{algorithmic}
\usepackage{graphicx}
\usepackage{textcomp}
\usepackage{xcolor}
\def\BibTeX{{\rm B\kern-.05em{\sc i\kern-.025em b}\kern-.08em
    T\kern-.1667em\lower.7ex\hbox{E}\kern-.125emX}}
\begin{document}

\title{CAMLPAD: Cybersecurity Autonomous Machine Learning Platform for Anomaly Detection\\}

\author{\IEEEauthorblockN{Ayush Hariharan}
\IEEEauthorblockA{\textit{Department of Computer Science} \\
\textit{Academy of Science}\\
Loudon County, USA \\
ahariharan@blue-cloak.com}
\and
\IEEEauthorblockN{Ankit Gupta}
\IEEEauthorblockA{\textit{Department of Computer Science} \\
\textit{TJHSST}\\
Alexandria, USA \\
2020agupta1@tjhsst.edu}
\and
\IEEEauthorblockN{Trisha Pal}
\IEEEauthorblockA{\textit{Department of Computer Science} \\
\textit{TJHSST}\\
Alexandria, USA \\
2021tpal@tjhsst.edu}

}

\maketitle

\begin{abstract}
As machine learning and cybersecurity continue to explode in the context of the digital ecosystem, the complexity of cybersecurity data combined with complicated and evasive machine learning algorithms leads to vast difficulties in designing an end-to-end system for intelligent, automatic anomaly classification. On the other hand, traditional systems use elementary statistics techniques and are often inaccurate, leading to weak centralized data analysis platforms. In this paper, we propose a novel system that addresses these two problems, titled CAMLPAD, for Cybersecurity Autonomous Machine Learning Platform for Anomaly Detection. The CAMLPAD system’s streamlined, holistic approach begins with retrieving a multitude of different species of cybersecurity data in real-time using elasticsearch, then running several machine learning algorithms, namely Isolation Forest, Histogram-Based Outlier Score (HBOS), Cluster-Based Local Outlier Factor (CBLOF), and K-Means Clustering, to process the data. Next, the calculated anomalies are visualized using Kibana and are assigned an outlier score, which serves as an indicator for whether an alert should be sent to the system administrator that there are potential anomalies in the network. After comprehensive testing of our platform in a simulated environment, the CAMLPAD system achieved an adjusted rand score of 95 percent, exhibiting the reliable accuracy and precision of the system. All in all, the CAMLPAD system provides an accurate, streamlined approach to real-time cybersecurity anomaly detection, delivering a novel solution that has the potential to revolutionize the cybersecurity sector.
\end{abstract}


\section{Introduction}
In recent years, the importance of varying fields within computer science, particularly cybersecurity and machine learning, has skyrocketed. With new systems depending on intelligent tools that bring next-level computation and systems open to security breaches, the importance of the intersection of machine learning for analysis in cybersecurity data has flourished. However, the burdens of uneven cybersecurity data from a variety of different sources often makes development of a tool that effectively and accurately makes use of machine learning to improve cybersecurity data difficult. As a result, very few end-to-end systems that can automatically classify anomalies in data exist, let alone those that are accurate.

In this paper, we propose a novel, accurate system for real-time anomaly detection. We term this product CAMLPAD, or the Cybersecurity and Autonomous Machine Learning Platform for Anomaly Detection. By processing a plethora of different forms of cybersecurity data, such as YAF, BRO, SNORT, PCAP, and Cisco Meraki real-time using a variety of machine learning models, our system immediately determines if a particular environment is at immediate risk for a breach as represented by presence of anomalies. The specific machine learning algorithms utilized include Isolation Forest, Histogram-Based Outlier Detection, Cluster-Based Local Outlier Factor, Multivariate Gaussian, and K-Means Clustering. Once the data has been processed and anomalies have been calculated, the CAMLPAD system utilizes Kibana to visualize outlier data pulled from Elasticsearch and to gauge how high the outlier score is. Once a particular threshold has been reached for this outlier score, an automated alert is sent to the system administrator, who has the option to forward the alert to all of the employees in the company so they are aware that a cybersecurity breach has occurred. By implementing CAMLPAD as a running bash script, the CAMLPAD system immediately recognizes anomalies and sends alerts.

\subsection{Background}
Cybersecurity is the practice of defense of an organization’s network and data from potential attackers, that have unauthorized access to the particular network. One measure of determining to what extent a particular user has this type of unauthorized access is detecting anomalies in the network traffic data, specifically the data referred to earlier (e.g. BRO, YAF, PCAP, SNORT). A potential platform that would be able to detect anomalies would need to process this data in real time then use a model that uses past data to learn whether current data contains anomalies and thus, if the network has an intrusion.

Specific to the CAMLPAD system, there are several ideas and terminologies that would benefit the reader to have a background of. To begin with, there are a few pieces of cybersecurity data that the CAMLPAD system makes use of. YAF, or “Yet Another Flowmeter”, is a cybersecurity data type that processes PCAP data and exports these flows to IPFIX Collecting Process [11]. BRO is an open source framework that analyze network traffic and is used to detects anomalies in a network. SNORT, similarly, is a network intrusion detection system that helps detect emerging threats.  Meraki is a cloud-based centralized management service that contains a network and organization structure. This specifically is crucial as it further reveals the relationships between members of a particular organizations, which further assists the machine learning model in determining where a potential anomaly may be.

Machine Learning, or ML, is the subfield within the exploding field of Artificial Intelligence primarily concerned with having a computer or machine learn to make predictions from a set of previous data rather than be explicitly programmed. There are several algorithms, or methods, that facilitate this type of learning. Machine learning consists of two main categories: unsupervised and supervised. In supervised machine learning, the ML model already has data labeled, so calculating an accuracy is as simple as detecting whether the model has correctly predicted the labeled data. In unsupervised machine learning, the main type of ML to be referred to in this paper, the data has not been labeled, so alternative methods need to be used to evaluate performance. As will be discussed in further detail in the next section, the specific machine learning models used as part of the CAMLPAD System are Isolation Forest, Histogram Based outlier detection, Cluster Based Local Outlier Factor, and Principal Component Analysis.

\begin{center}
\includegraphics[width=8cm,height=8cm]{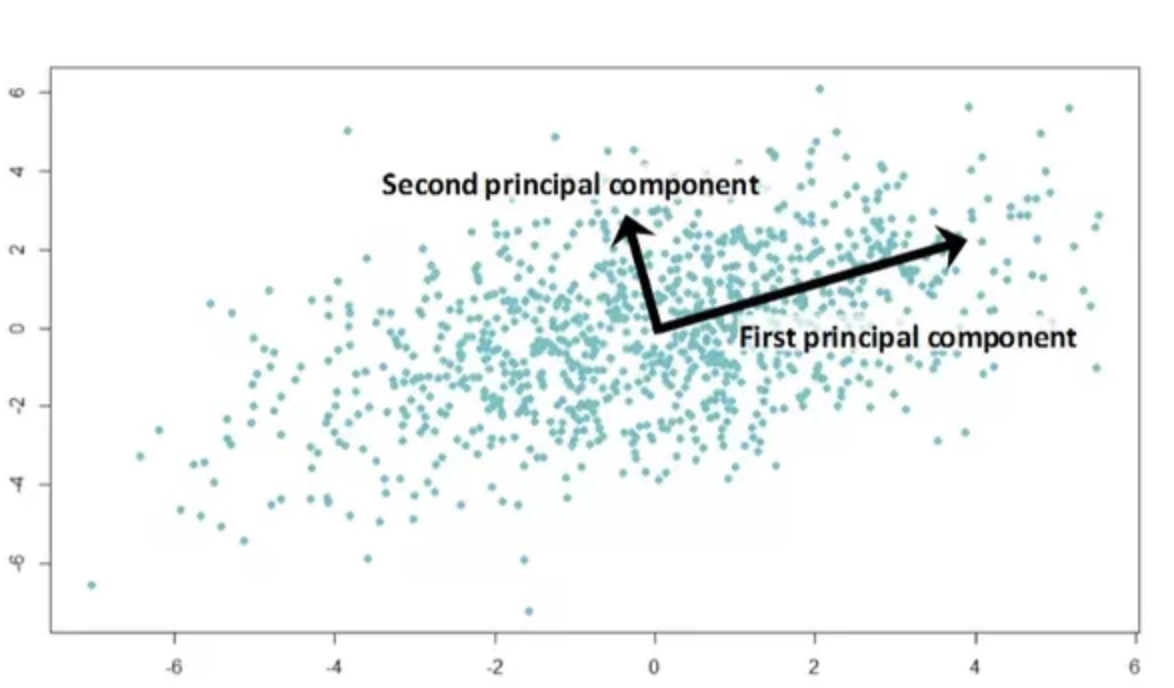}
\linebreak
Principal Component Analysis
\end{center}

Isolation forest is an unsupervised machine learning algorithm that randomly selects features by and selecting a value between the maximum and minimum for that selected feature. Since isolating normal points from anomalies requires more computation due to the need to cover a broader spectrum, an anomaly score can be utilized that measures the number of conditions needed to separate a given observation. The algorithm specifically begins by creating random decision trees, and then the score is calculated by being equal to the path length to isolate the observation. 

\begin{center}
\includegraphics[width=7cm,height=7cm]{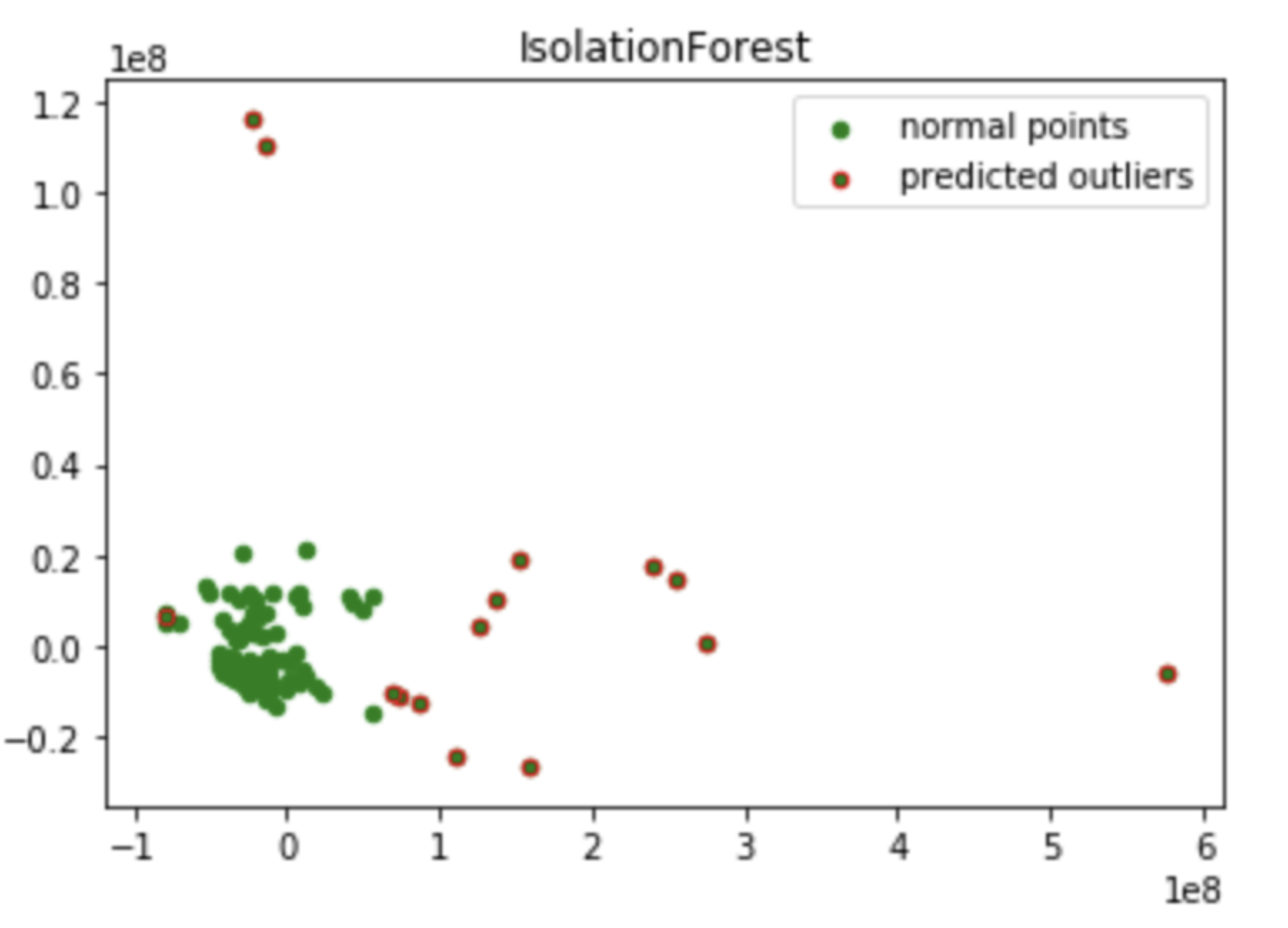}
\linebreak
Isolation Forest 
\end{center}

Histogram-based outlier detection is a method that scores records in linear time by assuming independence of features making it much faster than multivariate approaches, yet less precise. Cluster based local outlier factor (CBLOF) uses clusters to find anomalous data points by measuring local deviation of a given point with respect to its neighbors. Specifically, the CBLOF uses the concept of local density from k nearest neighbors by comparing densities of an object to those of its neighbors in order to identify regions of similar density.

\begin{center}
\includegraphics[width=8cm,height=6cm]{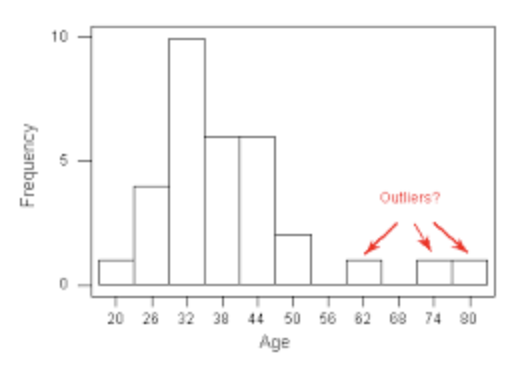}
Histogram-based Outlier Detection
\end{center}

Based off of the cybersecurity system, data types, and machine learning algorithms, the holistic CAMLPAD system incorporates the elasticsearch data stream in real time with the machine learning algorithm, and if the anomaly score reaches a particular threshold, a warning is sent to the organization. The methods section, delivers further details based off of the fundamentals discussed in this section.

\section{Methods}
	Before implementing the machine learning aspect of the research, the data, which described various online transactions that occurred at the Blue Cloak LLC headquarters,  had to be accurately transferred from the sensors, that were run on Linux virtual machines, to a local server, where the model can process the data and alert the user if any anomalies are present. The data, specific to the different sensors running on the virtual machine including BRO, YAF, Snort, and Meraki, are temporarily stored locally, in a machine composed of 4 Dell VRTX, before being uploaded to a hadoop server consisting of one master node and three slave nodes. After successfully uploading the data to the hadoop server, Apache NiFi is used to streamline and process the sensor logs before pushing the processed information into the Kafka Queue. Apache NiFi, a project from the Apache Software Foundation, was specifically designed to automate the flow of data between software systems. In our case, the data is transferred from the virtual storage on the hadoop server to the kafka queue where it can be more efficiently stores. Specifically, the information stored in each of the logs is queried into a json-like format consisting of a field, such as MAC address or destination ip, and the actual information, such as a list of actual addresses. Once the data has entered the queue, it is sent to the elasticsearch database where it is stored for future processing.
    \begin{center}
    \includegraphics[width=4cm,height=4cm]{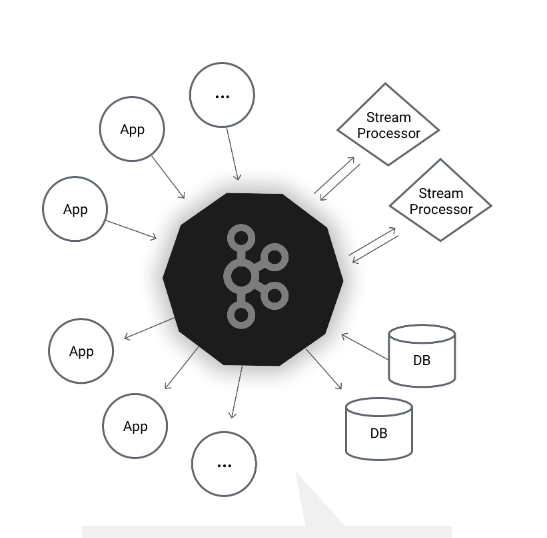}
\linebreak
Kafka Queue
\end{center}
	Elasticsearch is a database that parses and normalizes raw data before assigning each query of information a unique identification number. Using this identification number, and the index associated with the data, information from the sensor logs can be queried for further processing using the machine learning models. However, one caveat with Elasticsearch is that it doesn’t allow for custom processing scripts to be run inside the database. Instead, the information must be queried based on an assigned index and must be processed on an external server or node. That is where the current research enters the workflow, since the machine learning algorithms utilize the indexing ability of elasticsearch to stream data into a separate machine. This data is streamed directly from the database, without having to download the data as a CSV or JSON, meaning that the data is quickly transferred from storage to a local processor on another machine. Using a unique algorithm centered around the current date and time, all previous data is indexed and imported into a dataframe, one of the most common methods of storage for machine learning algorithms. The dataframe will contain the information used for training and validating the machine learning models that were created. Once the dataframe has been created, the current days data is indexed and imported into another dataframe. This dataframe will contain the latest information used for anomaly detection based upon patterns observed in the previous data stored. 

	Now that the data has been successfully imported into the respective dataframes, the categorical data present in the sensor logs, such as type of request or url, must be encoded into numerical values before further analysis by the machine learning algorithms. After encoding the data, two methods of imputation: linear regression, for purely numerical values, and backfill insertion, for encoded categorical values, were used. Now that missing or lost data has been imputed, the data can be imported into the custom ensemble model for anomaly detection. The custom ensemble model consists of an Isolation Forest algorithm, histogram based outlier detection algorithm, and cluster based local outlier factor. All of these models are similar to the implementation in the python outlier detection library. Once the data is fit to the overall model, the validation data and testing data are assigned an outlier score. Based on the outlier score and a simple PCA algorithm, clusters are developed depending on the outlier score assigned by the respective model. Those clusters are then processed and a heat map is created describing the various levels of outliers present in the data. 

	This process is repeated for each model created, resulting in three heat maps describing the outlier scores assigned by each model for the data. After the outlier scores have been assigned, the ensemble model is created through a democratic voting system, where each model has an equal say on whether a data point is an outlier or an inlier. After the voting system has been completed, the final outlier scores are run through the PCA algorithm and a final heat map is created. The process is then repeated for the different types of data that is stored on the Elasticsearch database, including YAF, BRO, SNORT, and Meraki. Specifically, the BRO data is split by protocol into DNS and CONN in order to accurately label the data. Once the final outlier scores have been compiled for each data type, a final ensemble model is created, using a democratic voting system in order to reclassify each data point. This final model takes into consideration not only different outlier detection models that have been successful in previous research, but also different types of sensor data, that capture different layers of internet traffic. After the model has been created, the accuracy is determined by calculating the Adjusted RAND score, a common method of evaluating unsupervised machine learning algorithms.

	This represents the last part of the workflow, where the data originated from different sensors has been effectively processed and anomaly scanning has been completed. After the accuracy has been tested and confirmed, the newly assigned outlier scores are then indexed and queried back into the Elasticsearch database so that visualizations of these outlier scores can be created in Kibana. Specifically, the index is portrayed as a gauge, where the outlier score of the current day’s data is compared with previous data that the model has trained on. When the gauge passes the 75th percentile, a custom alert is sent to the owner of the Apache database, alerting them that there might be anomalies in the current data. The user then has the choice of whether to respond to this alert by blocking certain destination ports or IP addresses or can perform further investigation to determine the cause of the anomaly.
	\begin{center}
\includegraphics[width=5cm,height=5cm]{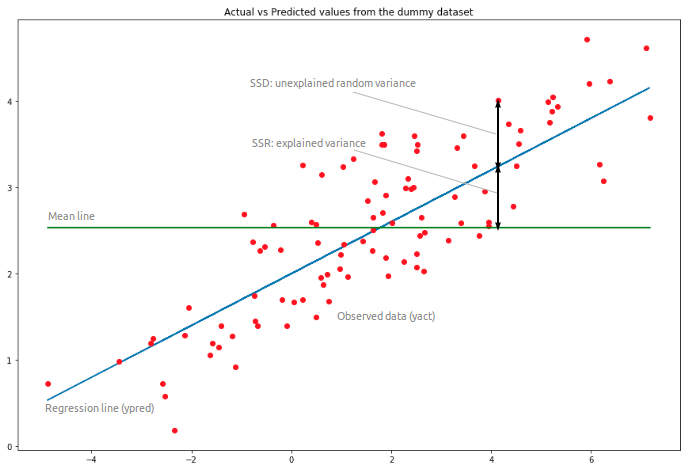}
\linebreak
Linear Regression
\end{center}

\section{Results}
In terms of results, the CAMLPAD system consists of five main data components: BRO DNS, BRO CONN, YAF, SNORT, and Meraki. These components, along with the three models: Isolation Forest (I-Forest), Cluster Based Local Outlier Factor (CBLOF), and Histogram Based Outlier Score (HBOS), are then combined in a democratic voting system to determine final outlier score. Although the user isn't alerted based upon each individual data type and only the final combined model, it is interesting to note how different types of data, both layer3 and layer2, show similar patterns in anomaly detection. In all of these heat maps, there are two seperate data points, previous data that the model trained on, which is represented by the smaller data points, and current days data, which is represented by the larger data points. In each graph, the differences or similarities between the current day and the previous data can be observed along with other patterns that represent the level of anomalies present.
\subsection{Data Heat Maps and Visualization}
As shown below, the BRO DNS data was processed by three main algorithms: HBOS, CBLOF, and I-Forest. In addition, the individual voting algorithm, isolated for the BRO DNS data, was implemented along with Kibana visualizations representing the recent day's data and previous days' data.
Based off of the HBOS heatmap, it is evident that, since many of the datapoints are lighter, there were more outliers, and the bigger dots for the current day were darker, meaning that the current day had less outliers. For the CBLOF heatmap, there were lighter dots at the center, surrounded by darker dots, meaning that the current day had more outliers than previous days. This means that the current day was likely an outlier. While the isolation forest algorithm does not display this, it IS further corroborated by the combined algorithm, which shows a dark dot representing less of an outlier, surrounded entirely by lighter dots. Since the Kibana Recent day has a score of 0.13, which is higher than 0.022 of the previous days, the current day is likely an outlier.
\begin{center}
\includegraphics[width=6cm,height=6cm]{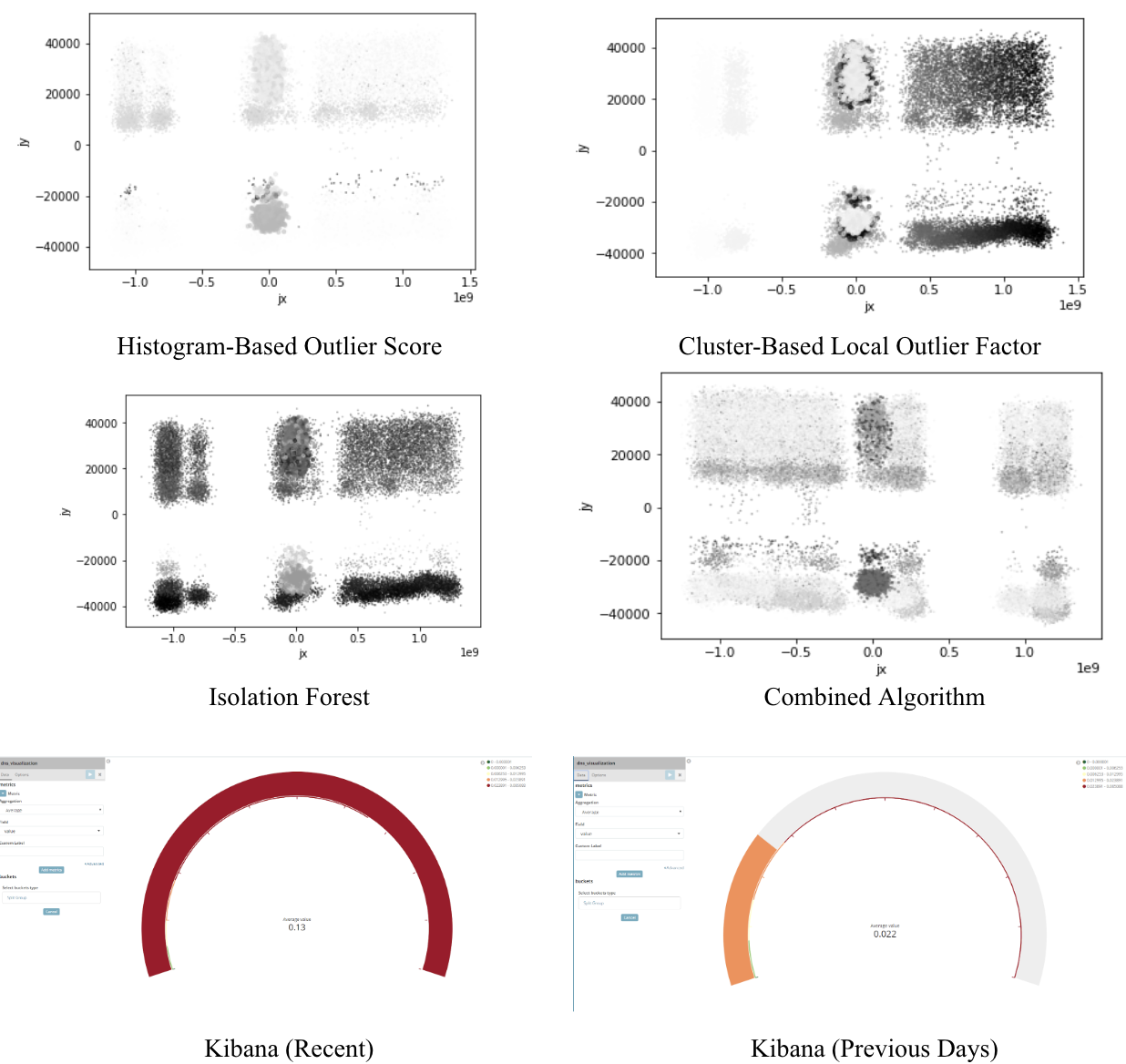}
\linebreak
BRO DNS Data Heat Maps and Kibana Visualization  
\end{center}
Next, the BRO CONN data's HBOS algorithm reveals that all of the data is very light, with a light gray dot in the center, meaning that the current day's data is only a little bit of an outlier in comparison tto previous days data. This is roughly corroborated by the CBLOF algorithm, but the I-Forest and Combined algorithms deliver an entirely different result. In both Isolation Forest and Combined, there is first a light dot surrounded by black dots, then a black dot surrounded by white dots, meaning that there was likely an outlier on the current day. Since the Kibana recent day has a score of 0.025, which is significantly lower than the previous days' score of 5.25, the Kibana score suggests that the current day is less of an anomaly.
\begin{center}
\includegraphics[width=6cm,height=6cm]{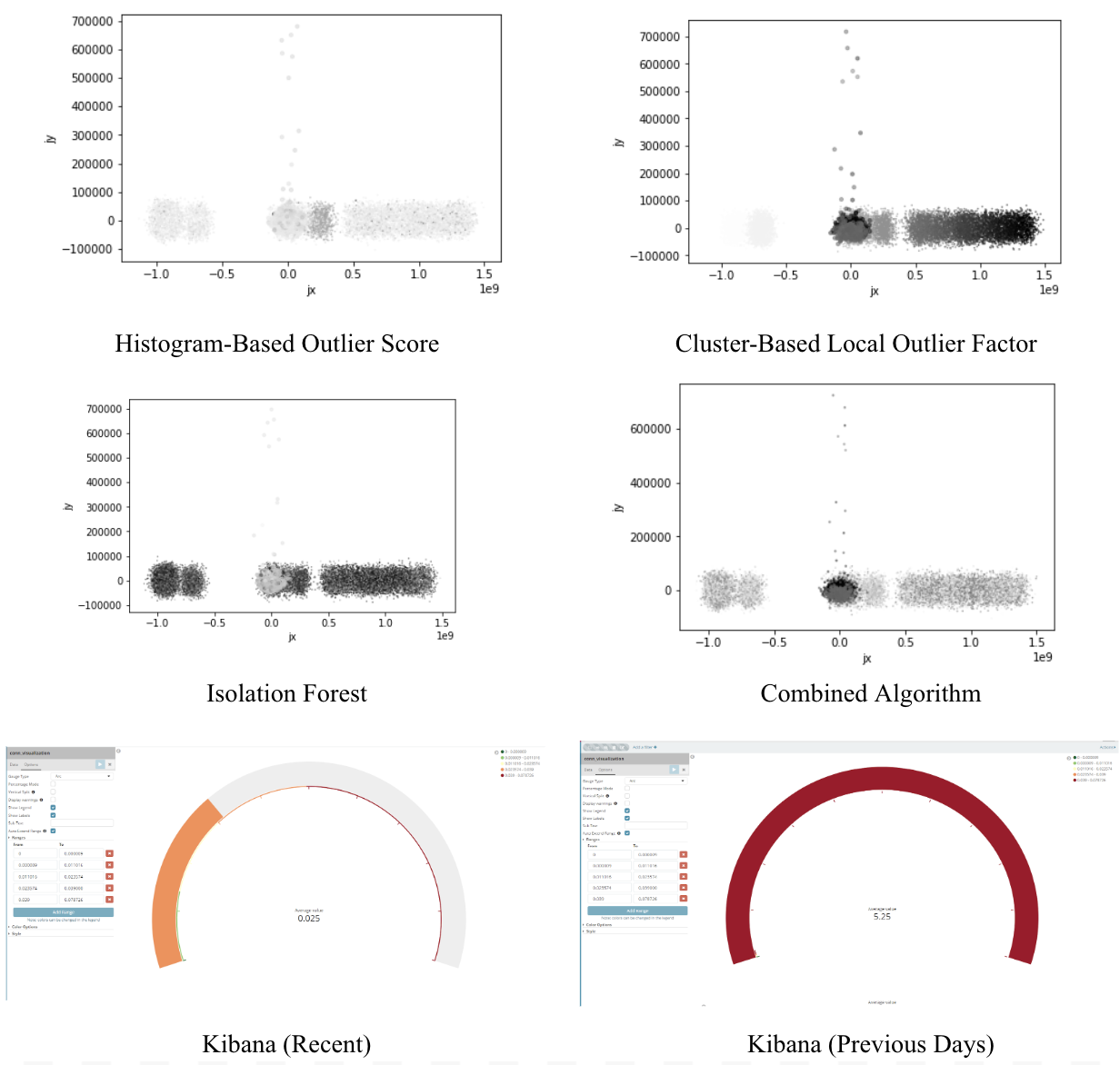}
\linebreak
BRO CONN Data Heat Maps and Kibana Visualization  
\end{center}
For the YAF data shown below, throughout each of the different algorithms, it is evident that the current day's data is an outlier when combined to the previous days' data. For example, in the HBOS diagram, there are black dots surrounded by light dots. This repeats for CBLOF, I-Forest, and the combined algorithm. For Kibana, the recent day has a score of 2.734, which is less than the previous days' score of 2.75, meaning that the most recent day is less of an outlier in comparison to previous days.

\begin{center}
\includegraphics[width=6cm,height=6cm]{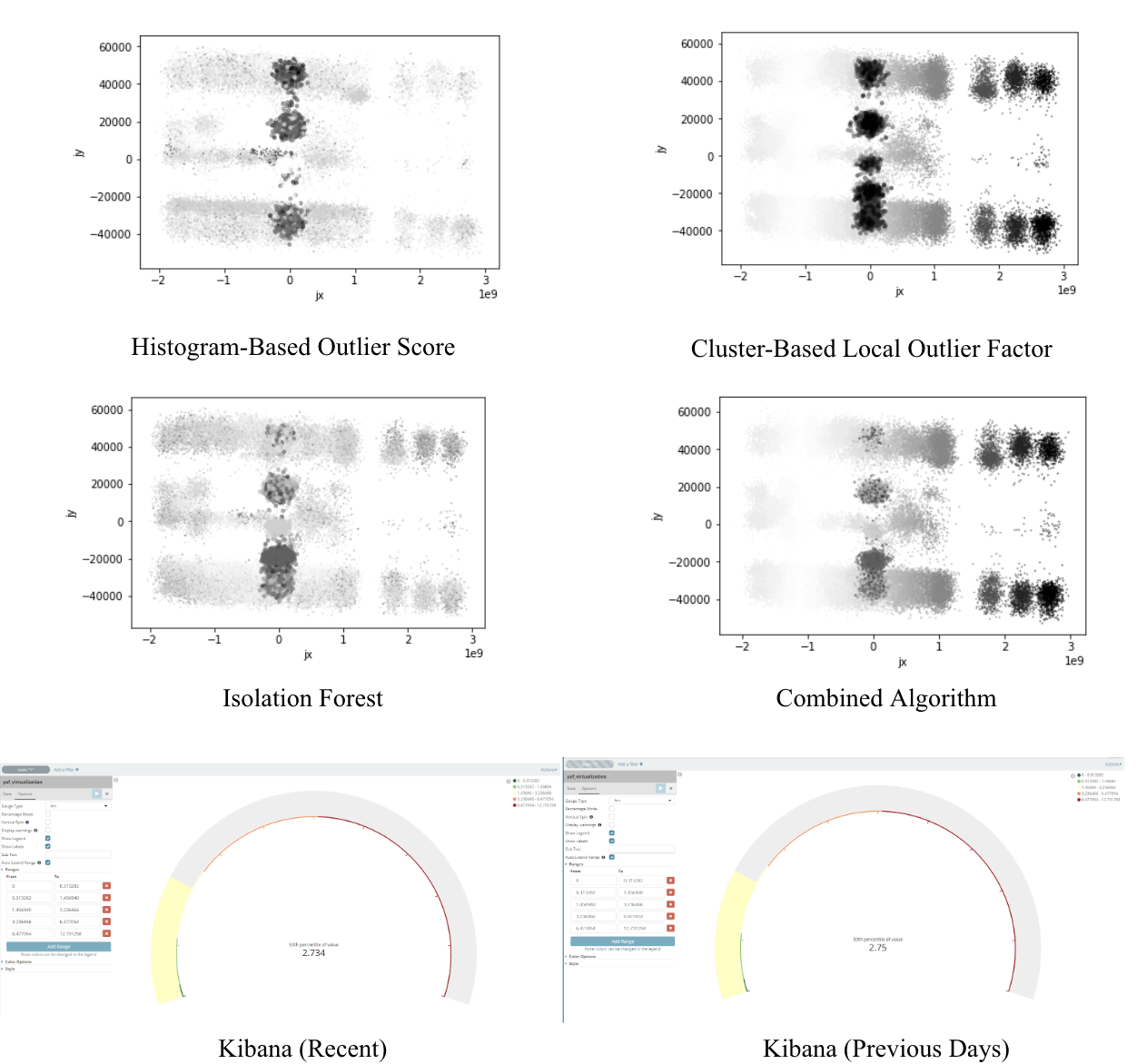}
\linebreak
YAF Data Heat Maps and Kibana Visualization  
\end{center}
For the SNORT data shown below, the HBOS, CBLOF, and Combined Algorithm all have black dots surrounded by lighter dots representing previous days' data. However, the I-Forest algorithm has only light dots, meaning that it does not perceive the current day's data to be an outlier.
For Kibana, the recent day has a score of 0.961, which is less than the previous days' score of 4.378, meaning that the most recent day is far less of an outlier in comparison to previous days.
\begin{center}
\includegraphics[width=8cm,height=8cm]{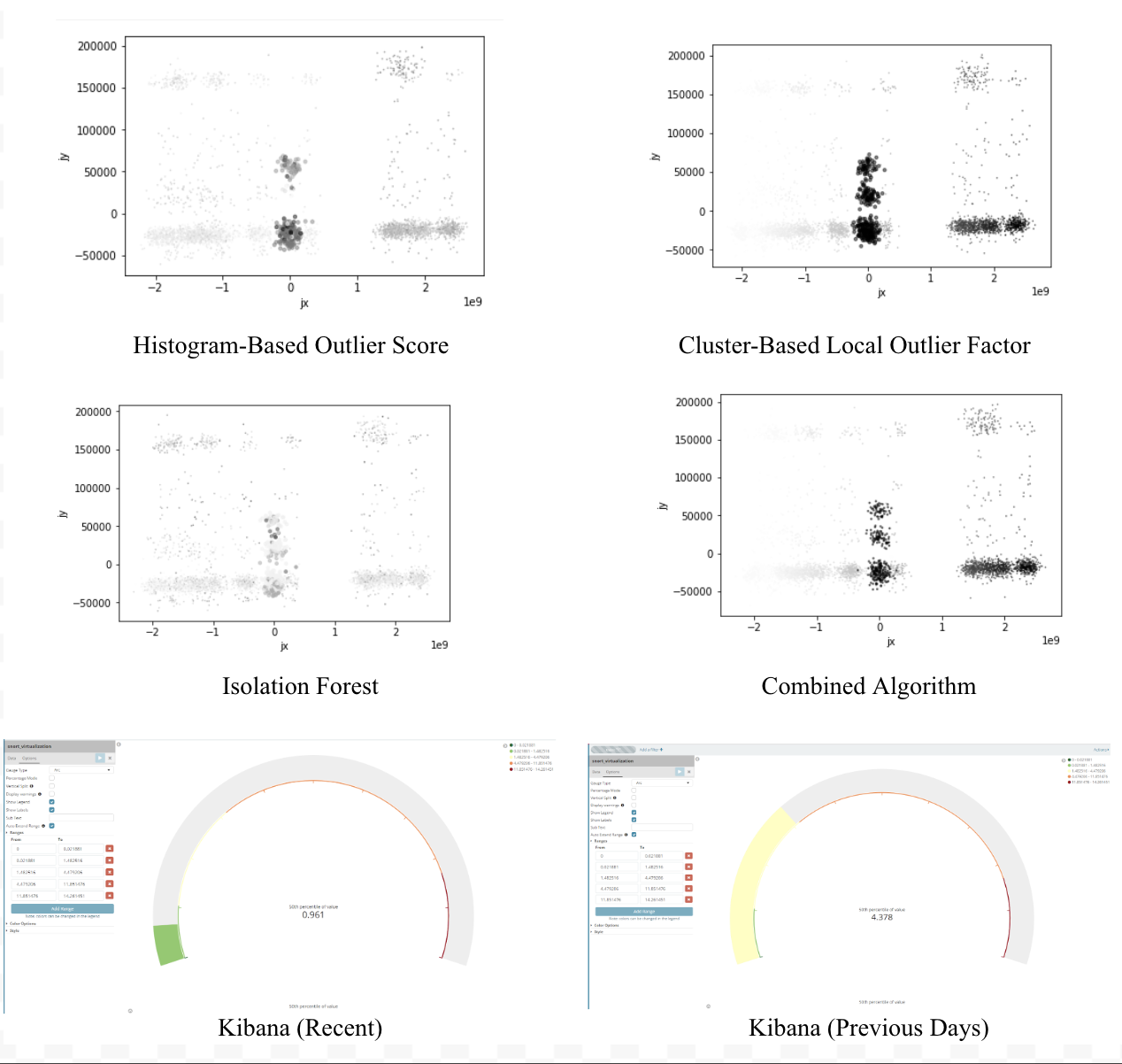}
SNORT Data Heat Maps and Kibana Visualization  
\end{center}
For the Meraki data shown below, it is clear that there are four main gray dots representing the current days' data, and these four clusters are outliers. While the HBOS model has the data being almost equivalent to the data from previous days, the CBLOF, I-Forest, and Combined all reveal that the clusters are indeed outliers. For Kibana, the recent day has a score of 5.287, which is less than the previous days' score of 0.029, meaning that the most recent day is far more of an outlier in comparison to previous days.
\begin{center}
\includegraphics[width=8cm,height=8cm]{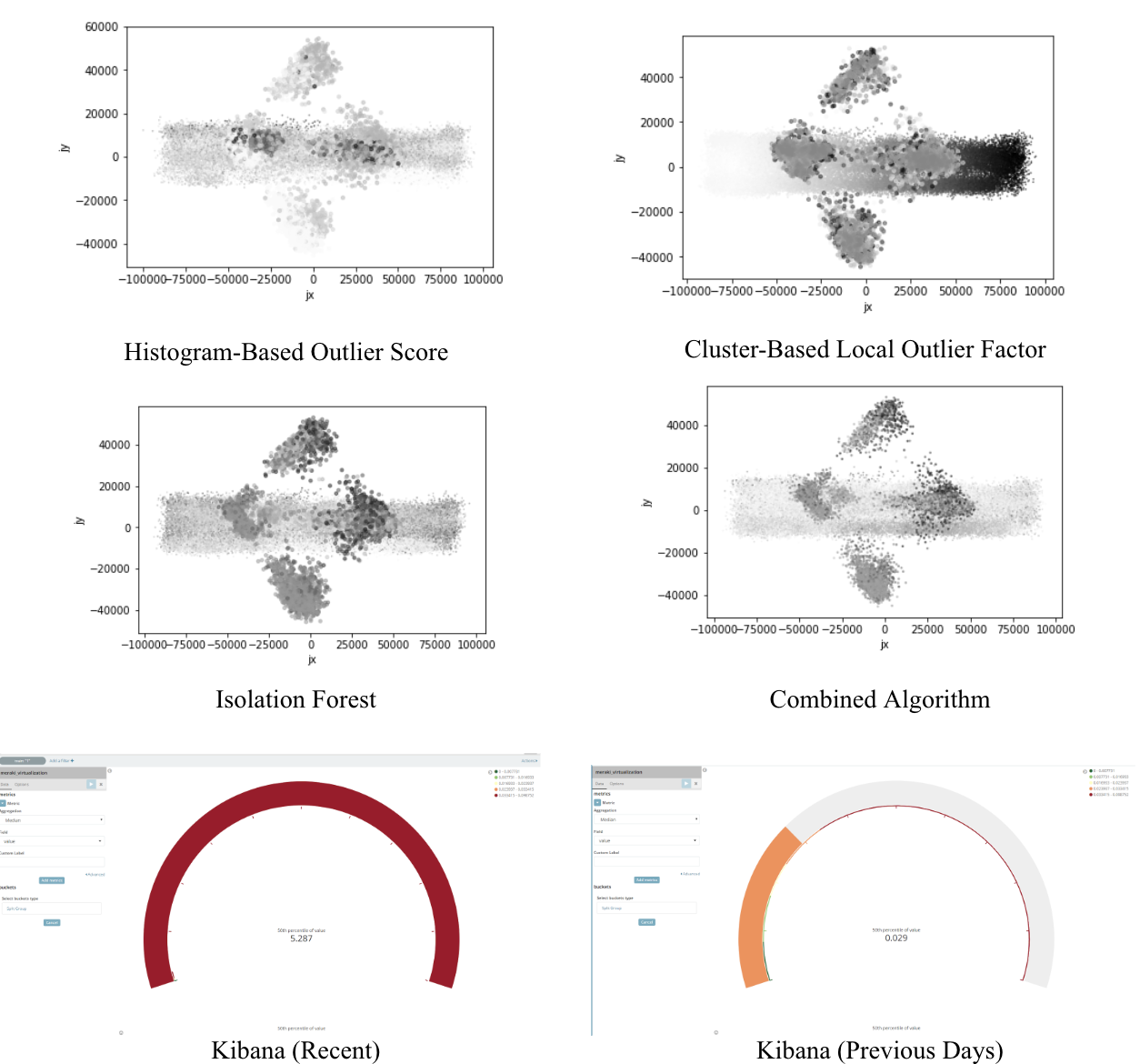}
Meraki Data Heat Maps and Kibana Visualization 
\end{center}
Lastly, for the combined data throughout each of the different algorithms, it is evident that the current day's data is an outlier when combined to the previous days' data. In HBOS, there is an X-like outlier, in the CBLOF there is a dark dot, in I-Forest, there is a gray dot, and in the combined algorithm there is a black dot surrounded by gray points. For Kibana, the recent day has a score of 0.261, which is more than the previous days' score of 0.004, meaning that the most recent day is more of an outlier in comparison to previous days.
\begin{center}
\includegraphics[width=8cm,height=8cm]{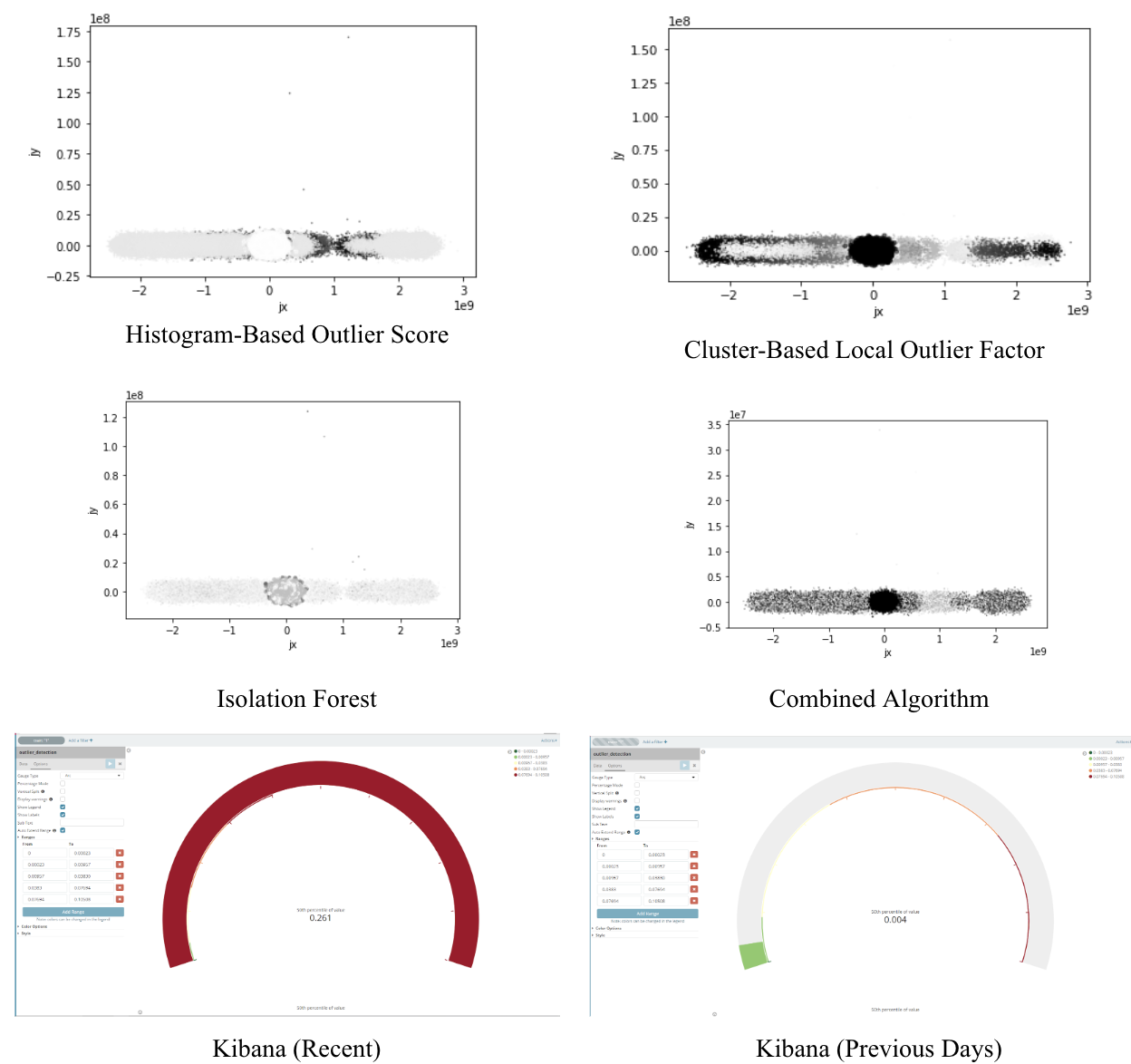}
Combined Data Heat Maps and Kibana Visualization 
\end{center}
\subsection{Accuracy Measure}
In order to determine our measure for accuracy, we used the RAND Score measure the similarity between two clusterings by considering all pairs of samples and counting pairs that are assigned. 
The RAND score takes into account True Positives, False Positives, and determines how accurate the measurement for a previous day is based off of the data from the current day. This is used, because a previous days' data is already validated, so it is easier to use the current day's data as a training set. All algorithms implemented achieve similar outlier scores, so it is evident that the holistic algorithm is correct, since none of the algorithms contradict each other. After testing with all algorithms implemented and averaging, we achieved a RAND score of 0.95.

\section{Discussion}
As previously discussed, the CAMLPAD System provides an innovative, comprehensive, and streamlined approach for anomaly detection. Not only does the system make use of a variety of data types, such as BRO (DNS/CONN), SNORT, YAF, and Meraki, it also uses a combined, democratic-voting based measure that incorporates all the different types of data to give a more holistic anomaly detection mechanism. Aside from the data, the use of four different algorithms, specifically Isolation Forest, Histogram Based Outlier Score, Cluster-Based Local Outlier Factor, and Angle-Based Outlier Detection, results in an integrated system that capitalizes on the strengths of each of the machine learning algorithms.

Although the CAMLPAD System has definitively proven to be accurate and useful due to the RAND score of 0.95, which corresponds to the two clusters being 95 percent similar to each other, it is also crucial to compare the system to previous network software architectures that have been traditionally used for anomaly detection.

Specifically, previous papers have reviewed different anomaly detection systems and found that there are specific issues with wide scale use deployment of intrusion detection systems [1]. Other papers have discussed the application of similar artificial intelligence-based algorithms with respect to the immune system, and how potentially there are interesting insights to be gained from computational models [2, 4]. Several researchers have also discussed more sophisticated machine learning algorithms, such as utilizing a spiking neural network algorithm, but even these have had limited success [3]. With the rise of machine learning, other researchers have investigated the subfield of deep learning, including its role as the frontier for distributed attack detection and in varied prevention systems [5, 6, 10, 17]. In addition, while several papers have pointed to the intriguing intersection of autonomous AI and anomaly detection, others contain a systematic review of how cloud computing could potentially assist in the development of a prevention system [7, 8, 16].

In addition to research papers that have investigated broad fields such as deep learning and cloud computing, several have dug deeper into the specifics of machine learning. For example, one paper considers use of a TCM-KNN algorithm for supervised network intrusion, which provides a novel approach, since most network intrusion algorithms use unsupervised learning [9]. In addition, another paper considers use of the n-gram machine learning algorithm in both anomaly detection and classification [18]. On the other hand, different papers have focused on larger scale intrusion networks, and more cost-benefit analysis oriented operations, involving research priorities and business intelligence [10, 12, 13, 15]. Real-time network flow analysis has been another field populated by a great deal of research, as shown by the system named Disclosure, which can detect botnet commands and control servers [14]. Lastly, several papers have investigated feature selection models and learning-based approaches to intrusion detection, such as applications to SQL attacks [19, 20].

\begin{center}
\includegraphics[width=2.5cm,height=2.5cm]{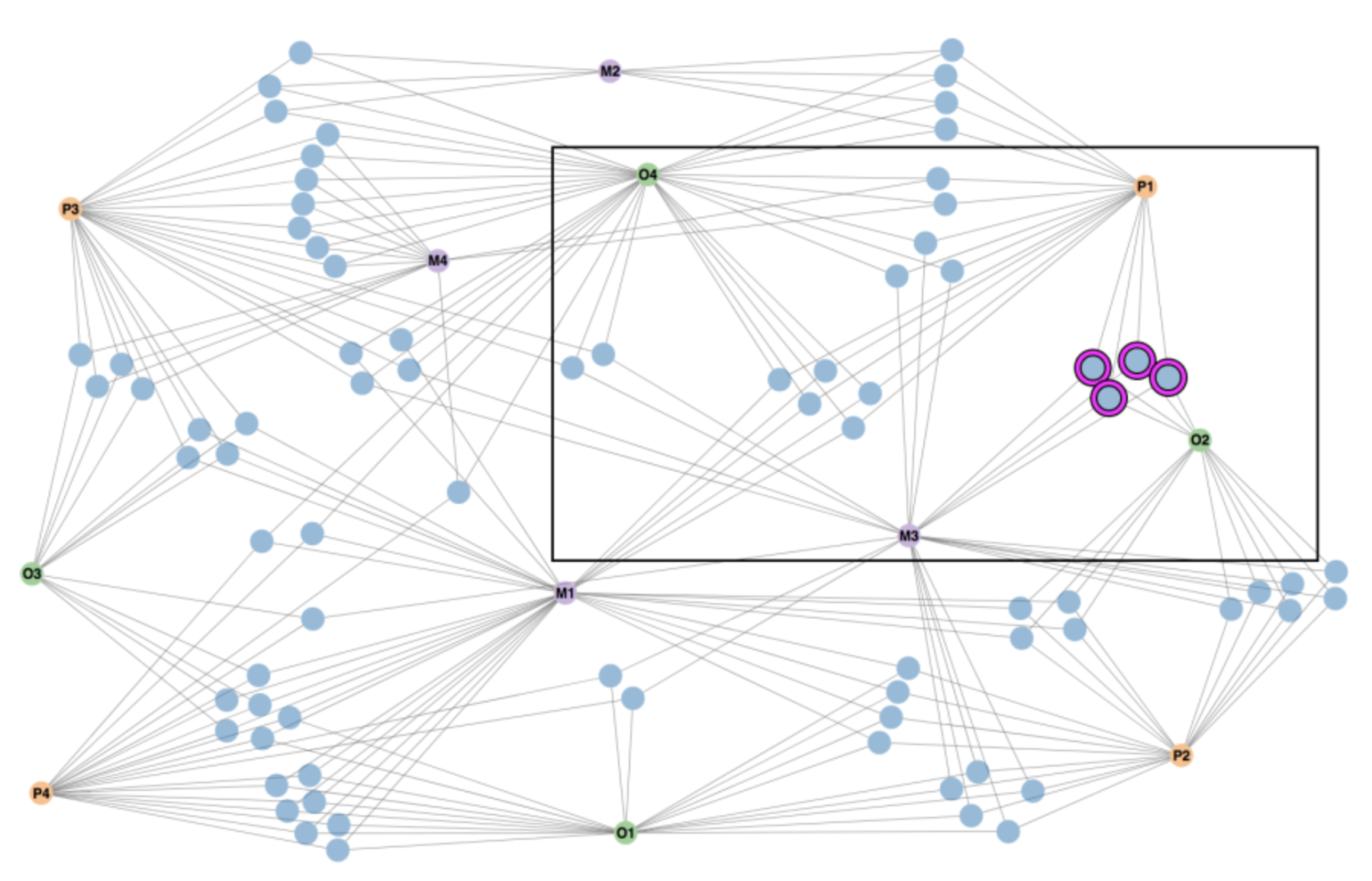}
\linebreak
Linear Regression
\end{center}

\begin{center}
\includegraphics[width=5cm,height=5cm]{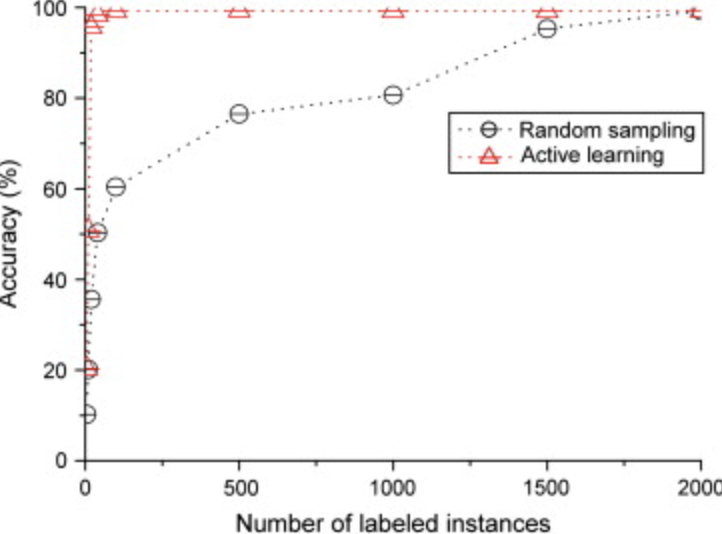}
\linebreak
TCM-KNN
\end{center}

As discussed, clearly a plethora of research has been done in the intersection of machine learning, cloud computing, and cybersecurity, but many research papers only include segments of a broader, holistic understanding of anomaly detection. Some only consider one type of data, while others only include one of a plethora of machine learning algorithms. In contrast, CAMLPAD addresses several types of data with the most efficient algorithms to provide a real-time, streamlined pipeline for autonomous anomaly detection.

\section{Conclusion}
Security breaches and threats are growing along with the cybersecurity field. CAMLPAD is machine learning-based platform that can effectively detect anomalies in real-time resulting in an outlier score. We demonstrated the vast possibilities of anomaly detection in the cybersecurity field using unsupervised machine learning models: Isolation Forest, Histogram-Based Outlier Detection, Cluster-Based Local Outlier Factor, Multivariate Gaussian, and K-Means Clustering.

CAMLPAD’s pipeline starts with data streamed directly from Elasticsearch and formatted on a local notebook. From there, we run the models which result in a visualization of the data and an outlier score. From the different models, the outlier scores are averaged and displayed on the Kibana dashboard.
\subsection{Future Research}
We plan to directly run our script on the The Metadata Encoding and Transmission Standard (METS) Schema to reach max efficiency and timely results. METS Schema helps encode descriptive and structural metadata for objects in a digital library. In addition to running the machine learning models on a METS Server, we plan to clean up the data and make each field consistent. Since the data, streamed from elasticsearch, is different for each group (based on day), preprocessing the data to make all data points similar will help increase the efficiency.

In the future, we will use supervised machine learning models to ensure all data points are represented. For example, we plan to use a Support Vector Machine (SVM) which is helpful for classification of the outliers and anomalies. The overall model is trained on data from an earlier period of time, then outliers are detected and scored based on those inconsistencies. 

Overall, CAMLPAD achieved an adjusted rand score of 95 percent, but with the use of several ML models and making CAMLPAD more efficient, the accuracy for anomaly detection can increase.

\section{Acknowledgements}
We would like to thank the employees at Blue Cloak, LLC for their generous support throughout the duration of this research endeavor as well as for the cybersecurity data and tools used.
\section*{References}
[1] Garcia-Teodoro, P., Diaz-Verdejo, J., Maciá-Fernández, G., Vázquez, E. (2009). Anomaly-based network intrusion detection: Techniques, systems and challenges. computers and security, 28(1-2), 18-28.

[2] Dasgupta, D. (Ed.). (2012). Artificial immune systems and their applications. Springer Science and Business Media.

[3] Demertzis, K., Iliadis, L., Spartalis, S. (2017, August). A spiking one-class anomaly detection framework for cyber-security on industrial control systems. In International Conference on Engineering Applications of Neural Networks(pp. 122-134). Springer, Cham.

[4] Dasgupta, D. (1999, October). Immunity-based intrusion detection system: A general framework. In Proc. of the 22nd NISSC (Vol. 1, pp. 147-160).

[5] Abeshu, A., Chilamkurti, N. (2018). Deep learning: the frontier for distributed attack detection in fog-to-things computing. IEEE Communications Magazine, 56(2), 169-175.

[6] Patel, A., Qassim, Q., Wills, C. (2010). A survey of intrusion detection and prevention systems. Information Management and Computer Security, 18(4), 277-290.

[7] Mylrea, M., Gourisetti, S. N. G. (2017). Cybersecurity and Optimization in Smart “Autonomous” Buildings. In Autonomy and Artificial Intelligence: A Threat or Savior? (pp. 263-294). Springer, Cham.

[8] Patel, A., Taghavi, M., Bakhtiyari, K., Junior, J. C. (2013). An intrusion detection and prevention system in cloud computing: A systematic review. Journal of network and computer applications, 36(1), 25-41.

[9] Li, Y., Guo, L. (2007). An active learning based TCM-KNN algorithm for supervised network intrusion detection. Computers and security, 26(7-8), 459-467.

[10] Diro, A. A., Chilamkurti, N. (2018). Distributed attack detection scheme using deep learning approach for Internet of Things. Future Generation Computer Systems, 82, 761-768.

[11] Inacio, C. M., Trammell, B. (2010, November). Yaf: yet another flowmeter. In Proceedings of LISA10: 24th Large Installation System Administration Conference (p. 107).

[12] Huang, M. Y., Jasper, R. J., Wicks, T. M. (1999). A large scale distributed intrusion detection framework based on attack strategy analysis. Computer Networks, 31(23-24), 2465-2475.

[13] Russell, S., Dewey, D., Tegmark, M. (2015). Research priorities for robust and beneficial artificial intelligence. Ai Magazine, 36(4), 105-114.

[14] Bilge, L., Balzarotti, D., Robertson, W., Kirda, E., Kruegel, C. (2012, December). Disclosure: detecting botnet command and control servers through large-scale netflow analysis. In Proceedings of the 28th Annual Computer Security Applications Conference (pp. 129-138). ACM.

[15] Chen, H., Chiang, R. H., Storey, V. C. (2012). Business intelligence and analytics: From big data to big impact. MIS quarterly, 36(4).

[16] Doelitzscher, F., Reich, C., Knahl, M., Passfall, A., Clarke, N. (2012). An agent based business aware incident detection system for cloud environments. Journal of Cloud Computing: Advances, Systems and Applications, 1(1), 9.

[17] Ten, C. W., Hong, J., Liu, C. C. (2011). Anomaly detection for cybersecurity of the substations. IEEE Transactions on Smart Grid, 2(4), 865-873.

[18] Wressnegger, C., Schwenk, G., Arp, D., Rieck, K. (2013, November). A close look on n-grams in intrusion detection: anomaly detection vs. classification. In Proceedings of the 2013 ACM workshop on Artificial intelligence and security (pp. 67-76). ACM.

[19] Aljawarneh, S., Aldwairi, M., Yassein, M. B. (2018). Anomaly-based intrusion detection system through feature selection analysis and building hybrid efficient model. Journal of Computational Science, 25, 152-160.

[20] Valeur, F., Mutz, D., Vigna, G. (2005, July). A learning-based approach to the detection 
of SQL attacks. In International Conference on Detection of Intrusions and Malware, and Vulnerability Assessment (pp. 123-140). Springer, Berlin, Heidelberg.
\end{document}